\documentclass[]{acoust}

\usepackage{cite}
\usepackage{comment}

\usepackage[T1]{fontenc}
\usepackage[ruled,vlined]{algorithm2e}
\SetKwInput{KwInput}{Input}
\SetKwInput{KwOutput}{Output}
\usepackage{array}
\usepackage{mdwmath}
\usepackage{mdwtab}
\usepackage{url}
\usepackage{multirow}
\usepackage{xcolor}
\usepackage{graphicx}
\usepackage{float}

\newcounter{subfig}[figure]
\newcommand{\subcaption}[1]{\footnotesize{(\refstepcounter{subfig}\alph{subfig}) #1}}

\def\header{K. MATSUMOTO and K. YATABE: SUBBAND SPLITTING FOR DETERMINED BSS}
\title{Subband Splitting: Simple, Efficient and Effective Technique for Solving Block Permutation Problem in Determined Blind Source Separation
}

\author{Kazuki Matsumoto$^{1,}$\thanks{k\_m\_w\_314@akane.waseda.jp} and Kohei Yatabe$^{2,}$\thanks{yatabe@go.tuat.ac.jp}}

\inst{$^{1}$Department of Communications and Computer Engineering, Waseda University, \\
      $3$--$4$--$1$ Ohkubo, Shinjuku-ku, Tokyo, $169$--$8555$ Japan\\
      $^{2}$Department of Electrical Engineering and Computer Science, Tokyo University of Agriculture and Technology,
      $2$--$24$--$16$ Naka-cho, Koganei-shi, Tokyo, $184$--$8588$ Japan\\
      }

\abst{
Solving the permutation problem is essential for determined blind source separation (BSS).
Existing methods, such as independent vector analysis (IVA) and independent low-rank matrix analysis (ILRMA), tackle the permutation problem by modeling the co-occurrence of the frequency components of source signals.
One of the remaining challenges in these methods is the block permutation problem, which may cause severe performance degradation.
In this paper, we propose a simple and effective technique for solving the block permutation problem. 
The proposed technique splits the entire frequency bands into several overlapping subbands and sequentially applies BSS methods (e.g., IVA, ILRMA, or any other method) to each subband.
Since the splitting reduces the size of the problem, the BSS methods can effectively work in each subband.
Then, the permutations among the subbands are aligned by using the separation result in one subband as the initial values for the other subbands.
Additionally, we propose SS-IVA and SS-ILRMA by combining subband splitting (SS) with IVA and ILRMA.
Experimental results demonstrated that our technique remarkably improves the separation performance without increasing computational cost.
In particular, our SS-ILRMA achieved the separation performance comparable to the oracle method (frequency-domain independent component analysis with the ideal permutation solver).
Moreover, SS-ILRMA converged faster than conventional IVA and ILRMA.
}

\kword{
Multichannel source separation, 
independent vector analysis (IVA), 
independent low-rank matrix analysis (ILRMA),
optimization, 
band splitting.
}

\makeatletter
\long\def\@makefntext#1{\vskip3\p@ \hsize\columnwidth \par \noindent \footnotesize \hskip6\p@ $^{\@thefnmark}$\hskip1\p@#1\vskip-3\p@}
\makeatother

\begin{document}
\maketitle

\section{Introduction}

Determined blind source separation (BSS) is 
a technique for separating the source signals from multichannel observed signals. 
It is usually formulated as an optimization problem of the demixing matrices in the time-frequency domain.
To separate the signals with this formulation, addressing the \textit{permutation problem} is essential.
Namely, the order of extraction target must be kept consistent across all frequencies\cite{smaragdisBlindSeparationConvolved1998,murataApproachBlindSource2001a}.
While a permutation solver\cite{sawadaRobustPreciseMethod2004,wangRegionGrowingPermutationAlignment2011,sarmientoContrastFunctionBased2015,yamajiDNNBasedPermutationSolver,hasuikeDNNBasedFrequencyDomainPermutation2022,liHBPEfficientBlock2022,emuraPermutationAlignmentMethodUsing2024} is required in frequency-domain independent component analysis (FDICA)\cite{smaragdisBlindSeparationConvolved1998,murataApproachBlindSource2001a}, later methods, e.g., independent vector analysis (IVA)\cite{hiroeSolutionPermutationProblem2006,kimIndependentVectorAnalysis2006,onoStableFastUpdate2011,scheiblerFastStableBlind2020,yatabeTimefrequencymaskingbasedDeterminedBSS2019,yatabeDeterminedBSSBased2021,liangOvercomingBlockPermutation2012,jangIndependentVectorAnalysis2007,leeIndependentVectorAnalysis2012, ikeshitaIndependentVectorAnalysis2017, shinAuxiliaryFunctionBasedIndependentVector2020,liGeometricallyConstrainedIndependent2020, gotoGeometricallyConstrainedIndependent2022}, independent low-rank matrix analysis (ILRMA)\cite{kitamuraDeterminedBlindSource2016,sawadaReviewBlindSource2019,kitamuraConsistentIndependentLowrank2020,mogamiIndependentLowrankMatrix2017,mogamiIndependentLowRankMatrix2019,mitsuiVectorwiseCoordinateDescent2018,oshimaInteractiveSpeechSource2021,ikeshitaIndependentLowRankMatrix2018}, and those assisted by deep learning \cite{makishimaIndependentDeeplyLearned2019,hasumiEmpiricalBayesianIndependent2021,hasumiPoPIDLMAProductofPriorIndependent2023,kameokaSupervisedDeterminedSource2019a,liFastMVAEJoint2019a,liFastMVAE2ImprovingAccelerating2023,scheiblerSurrogateSourceModel2021,matsumotoDeterminedBSSCombination2024}, model the co-occurrence of the frequency components and align the permutation within the optimization algorithms.

\begin{figure*}[t]
    \centering
    \includegraphics[width = \linewidth]{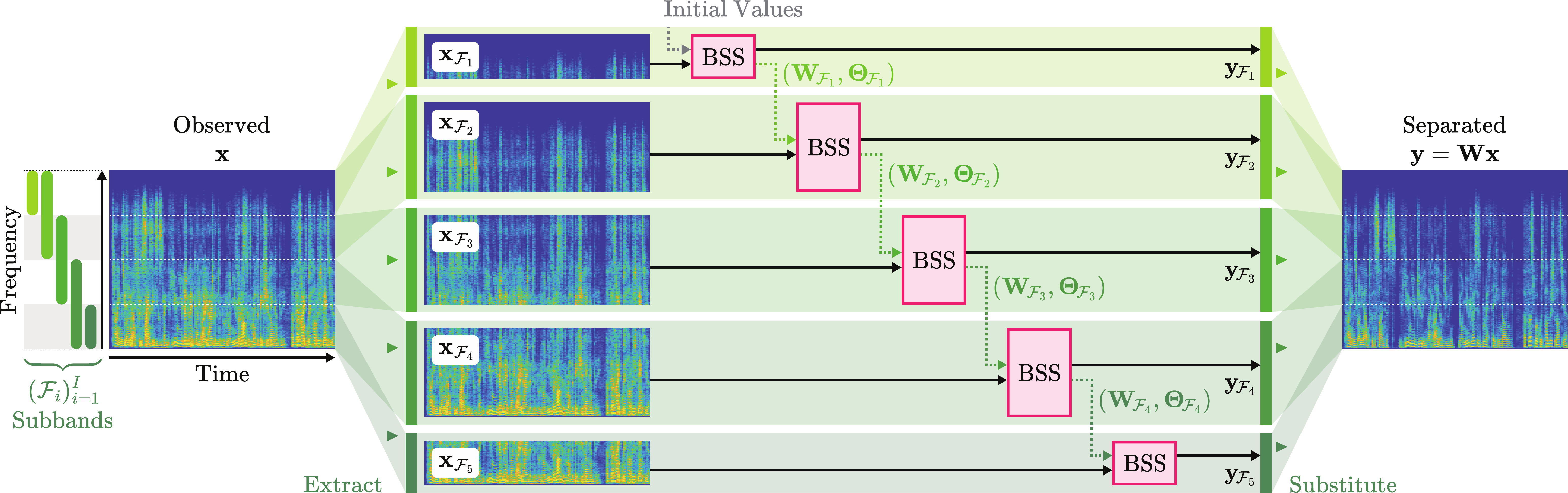}
    \caption{Illustration of the proposed technique named \textit{subband splitting}. 
    The observed signal $\mathbf{x}$ is split into overlapping subbands 
    $(\mathbf{x}_{\mathcal{F}_i})_{i=1}^I$, and a BSS method (e.g., IVA, ILRMA, or any other method) sequentially separates each subband 
    ${\mathcal{F}_i}$ by using the demixing matrices $\mathbf{W}_{\!\mathcal{F}_i}$.
    The separation results in the $i$th subband $\mathcal{F}_{i}$, including the auxiliary variable $\boldsymbol{\Theta}_{\!\mathcal{F}_i}$, 
    is used as the initial values in the next subband $\mathcal{F}_{i+1}$, which aligns the permutation among the subbands.}
    \label{fig:concept}
\end{figure*}

Although these methods can align the permutation to some extent, they sometimes fail due to \textit{block permutation problem}\cite{liangOvercomingBlockPermutation2012,mitsuiVectorwiseCoordinateDescent2018,oshimaInteractiveSpeechSource2021,liHBPEfficientBlock2022,liFastMVAE2ImprovingAccelerating2023}.
That is, several frequency blocks with inconsistent permutations may arise and cause severe performance degradation.
To deal with the block permutation problem, the following three types of approaches have been employed in many existing methods.
The first one utilizes an external permutation solver tailored for block permutation problem\cite{oshimaInteractiveSpeechSource2021,liHBPEfficientBlock2022}. The second one incorporates spatial information (e.g., the directions of the sources) into BSS algorithms\cite{mitsuiVectorwiseCoordinateDescent2018, liGeometricallyConstrainedIndependent2020, gotoGeometricallyConstrainedIndependent2022}.
The last approach is to improve the source models.
In particular, aiming to precisely model the frequency-band-wise structures of the sources, several methods incorporated subband structures into the source model of IVA and ILRMA\cite{liangOvercomingBlockPermutation2012,jangIndependentVectorAnalysis2007,leeIndependentVectorAnalysis2012,  ikeshitaIndependentVectorAnalysis2017, shinAuxiliaryFunctionBasedIndependentVector2020, ikeshitaIndependentLowRankMatrix2018}.
These three approaches have mitigated the block permutation problem.
However, some of these methods may require additional computational costs for external solvers (the first approach), while others may require efforts to develop optimization algorithms, especially when incorporating more sophisticated source models (the second and third approaches).

In this paper, we propose a simple technique named \textit{subband splitting} (SS) to enhance the separation performance of existing BSS algorithms\footnote{
The same proposal has already been uploaded on arXiv\cite{matsumoto2025subbandsplittingsimpleefficient}
}.
As illustrated in Fig. \ref {fig:concept}, our technique splits the entire frequency bands into several overlapping subbands and then sequentially applies BSS algorithms to each subband.
Owing to the splitting, the size of the optimization problem in each subband is reduced, and therefore source separation can be more easily done.
Then, the separation results from one subband are used to initialize BSS algorithms in the subsequent subbands, which aligns the permutations across all subbands.
Notably, our technique does not require any modification to the BSS algorithms nor additional computational cost.
Therefore, it can be directly combined with various BSS algorithms and can improve their separation performance without paying additional costs.

We additionally propose SS-IVA and SS-ILRMA, in which IVA and ILRMA are sequentially applied to each subband.
They demonstrate the usefulness of our technique.
The experimental results showed that our technique improved the separation performance of both IVA and ILRMA.
The robustness of ILRMA was especially improved, achieving a separation performance comparable to the oracle method (FDICA with the ideal permutation solver (IPS)).
Moreover, the proposed SS-IVA and SS-ILRMA empirically required fewer iterations and shorter runtime to converge, compared to the conventional IVA and ILRMA.

The rest of the paper is organized as follows. 
Section \ref{sec:prelimi} outlines determined BSS and the block permutation problem. 
In Section \ref{sec:prop}, we propose the subband splitting technique for arbitrary BSS algorithms.
In Section \ref{sec:exp}, we propose SS-IVA and SS-ILRMA and experimentally investigate their separation performance and runtime.
Section \ref{sec:conclusion} concludes the paper.

\begin{figure*}[t]
\begin{minipage}[t]{0.3\linewidth}
    \centering
    \includegraphics[scale=0.32]{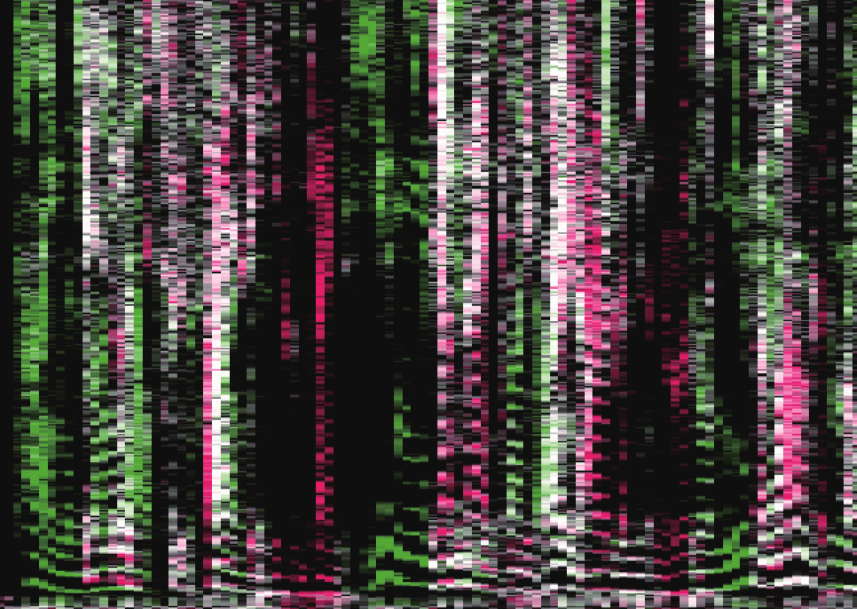}\\
    \subcaption{Observed}
\end{minipage}
\hfill
\begin{minipage}[t]{0.3\linewidth}
    \centering
    \includegraphics[scale=0.32]{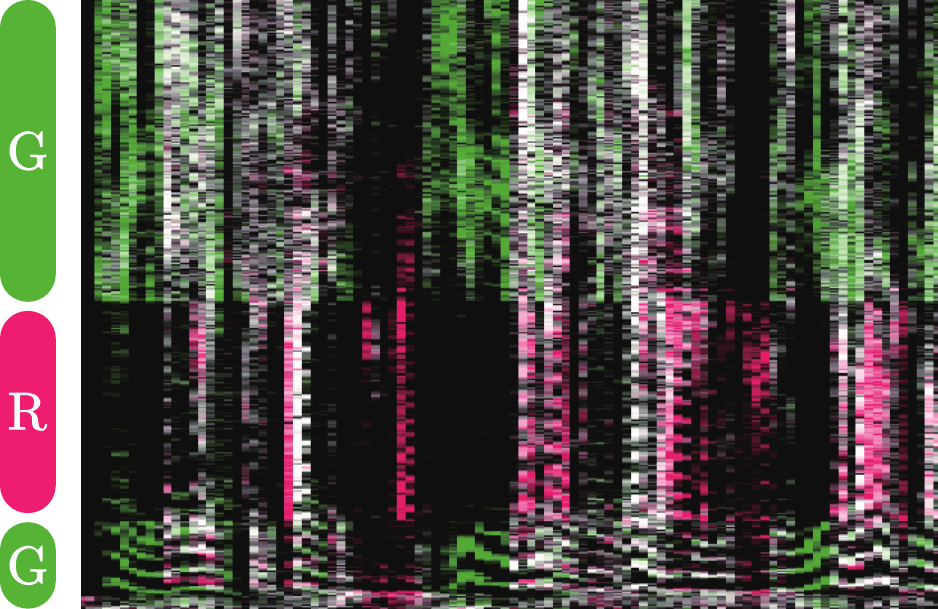}\\
    \subcaption{IVA (SDRi = 1.29 dB)}
\end{minipage}
\hfill
\begin{minipage}[t]{0.3\linewidth}
    \centering
    \includegraphics[scale=0.32]{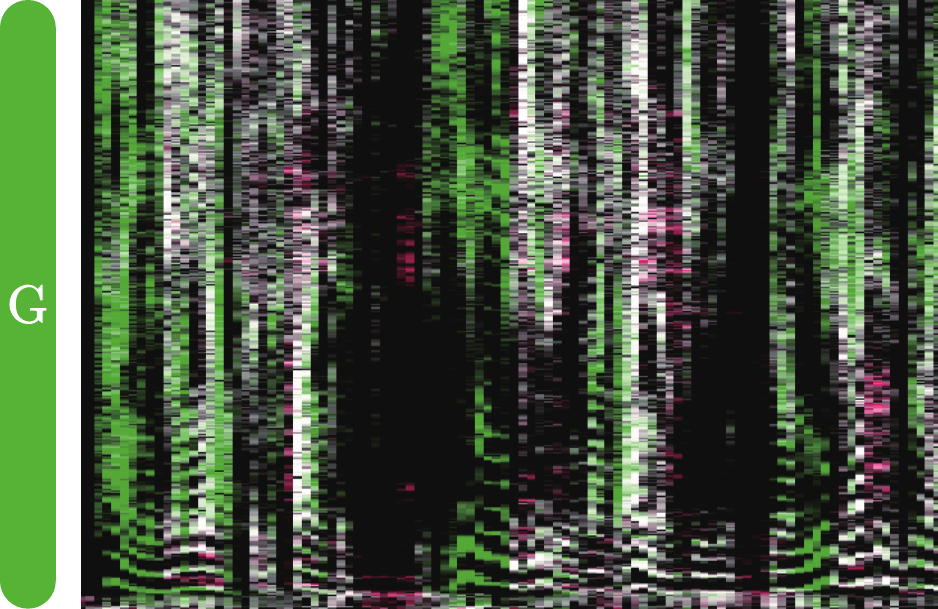}\\
    \subcaption{Our SS-IVA (SDRi = 14.44 dB)}
\end{minipage}
\caption{
Example of separation results and their evaluation metric (SDRi) in a two-channel and two-source situation ($M=N=2$).
The observed signal (a) is generated from \texttt{dev1\_male4\_src\_1.wav} and \texttt{dev1\_male4\_src\_2.wav} in SiSEC 2011 dataset\cite{araki2011SignalSeparation2012}, by convolving them with room impulse responses recorded in a real room \cite{hadadMultichannelAudioDatabase2014}.
The sources are placed at $-75^\circ$ and $60^\circ$, respectively, and the other experimental conditions are the same as those described in Section \ref{sec:commonCond}
The green and red colors represent the two different sources, where the ratio of their energy was calculated using the oracle sources.
For visibility, the frequency axis is trimmed from 0 to 3\,kHz.
For the separated signals (b) and (c), the colors of the left bars indicate the dominant source in each frequency band, where the letter G and R corresponds to green and red, respectively.
While the conventional IVA in (b) resulted in poor SDRi due to the block permutation problem, our proposed SS-IVA was able to successfully align the permutations across all frequencies, even though the BSS algorithm used in (b) and (c) was the same (i.e., AuxIVA\cite{onoStableFastUpdate2011}).
}
\label{fig:perm}
\end{figure*}

\section{Preliminaries}
\label{sec:prelimi}

\subsection{Determined BSS}

Determined BSS can be formulated as an optimization problem of the demixing matrices in the time-frequency domain.
Let $\mathbf{s}_{ft}=[s_{ft1},\ldots, s_{ftN}]^\mathsf{T}\in\mathbb{C}^{N}$ be a 
vector of $N$ source signals at the $(f,t)$th bin, where $1\leq f\leq F$ and $1\leq t\leq T$ are frequency and time indices, respectively.
The $M$-channel observed signal $\mathbf{x}_{ft}=[x_{ft1},\ldots, x_{ftM}]^\mathsf{T}\in\mathbb{C}^{M}$ is approximated using the frequency-wise mixing matrix $\mathbf{A}_f\in\mathbb{C}^{M\times N}$ as $\mathbf{x}_{ft} = \mathbf{A}_f\mathbf{s}_{ft}$.
In a determined situation (i.e., $N \leq M$), the $n$th separated signal at the $(f,t)$th bin is extracted using the $n$th demixing vector $\mathbf{w}_{\!fn}\in\mathbb{C}^M$ as follows:
\begin{equation}
    y_{ftn} = \mathbf{w}_{\!fn}^\mathsf{H}\mathbf{x}_{\!ft}.
    \label{eq:prodSeparation}
\end{equation}
Then, the $N$ separated signals $\mathbf{y}_{\!ft}=[y_{ft1},\ldots, y_{ftN}]^\mathsf{T}\in\mathbb{C}^{N}$ are obtained by\begin{equation}
    \mathbf{y}_{\!ft} = \begin{pmatrix}
        y_{ft1}\\
        \vdots\vspace{-3pt}\\
        y_{ftN}\\
    \end{pmatrix} = \begin{pmatrix}
        \mathbf{w}_{\!f1}^\mathsf{H}\mathbf{x}_{\!ft}\\
        \vdots\vspace{-3pt}\\
        \mathbf{w}_{\!fN}^\mathsf{H}\mathbf{x}_{\!ft}\\
    \end{pmatrix} = \mathbf{W}_{\!f}\mathbf{x}_{ft},
    \label{eq:LinearSeparation}
\end{equation}
where
\begin{equation}
    \mathbf{W}_{\!f}=\begin{pmatrix}
        \mathbf{w}_{\!f1}^\mathsf{H}\\
        \vdots\vspace{-3pt}\\
        \mathbf{w}_{\!fN}^\mathsf{H}
    \end{pmatrix}\in\mathbb{C}^{N\times M}
    \label{eq:matrix}
\end{equation}
is the demixing matrix.
For notational simplicity, we omit the indices to represent all the components altogether as $\mathbf{x}=((\mathbf{x}_{ft})_{f=1}^F)_{t=1}^T$, $\mathbf{y}=((\mathbf{y}_{\!ft})_{f=1}^F)_{t=1}^T$ and $\mathbf{W}=(\mathbf{W}_{\!f})_{f=1}^F$.
Then, the linear operation in Eq.~\eqref{eq:LinearSeparation} for all frequencies $f=1,\ldots,F$ and times $t=1,\ldots,T$ is shortly represented as $\mathbf{y}=\mathbf{W}\mathbf{x}$ for brevity.

To separate the sources using Eq.~\eqref{eq:LinearSeparation}, addressing the permutation problem is essential.
Namely, the extraction target of demixing vector $\mathbf{w}_{\!fn}^\mathsf{H}$ in Eq. \eqref{eq:prodSeparation} must be consistent across all frequencies $f =1,\ldots,F$.
A standard approach to the permutation problem is to model the co-occurrence of the frequency components.
For instance, IVA \cite{hiroeSolutionPermutationProblem2006,kimIndependentVectorAnalysis2006} considers the frequency-directional group structure, which results in the minimization problem of the following objective function:
\begin{equation}
   \mathcal{L}(\mathbf{W})= \sum_{n=1}^N\sum_{t=1}^T \sqrt{\sum_{f=1}^F|y_{ftn}|^2} - 2T\sum_{f=1}^F\log(|\det(\mathbf{W}_{\!f})|),
\label{eq:costIVA}
\end{equation}
where $y_{ftn} = \mathbf{w}_{\!fn}^\mathsf{H}\mathbf{x}_{\!ft}$.
ILRMA\cite{kitamuraDeterminedBlindSource2016} utilizes nonnegative matrix factorization (NMF) 
\cite{fevotteNonnegativeMatrixFactorization2009b}
to model low-rankness of the power spectrograms of sources.
The typical objective function is
\begin{align}
\mathcal{L}(\mathbf{W},\boldsymbol{\Theta})=& \sum_{n=1}^N\sum_{t=1}^T\sum_{f=1}^F\left(\frac{|y_{ftn}|^2}{\mathbf{t}_{fn}^\mathsf{T}\!\mathbf{v}_{tn}}+\log(\mathbf{t}_{fn}^\mathsf{T}\!\mathbf{v}_{tn})\right) \nonumber\\
&\hspace{50pt}-2T\sum_{f=1}^F\log(|\det(\mathbf{W}_{\!f})|),
\label{eq:costILRMA}
\end{align}
where $\boldsymbol{\Theta}=((\mathbf{T}_n)_{n=1}^N,(\mathbf{V}_n)_{n=1}^N)$ 
represents the set of auxiliary variables for NMF,  
$\mathbf{T}_n =[\mathbf{t}_{1n},\ldots,\mathbf{t}_{Fn}]^\mathsf{T}\in\mathbb{R}_+^{F\times K}$ and 
$\mathbf{V}_n = [\mathbf{v}_{1n},\ldots,\mathbf{v}_{Tn}]\in\mathbb{R}_+^{K \times T}$
are the basis and activation matrices, respectively, and $K$ is the number of bases.

\subsection{Block Permutation Problem}
\label{sec:blockPermutation}

Despite the great success of the above methods, they occasionally fail in separation due to the block permutation problem\cite{liangOvercomingBlockPermutation2012,mitsuiVectorwiseCoordinateDescent2018,oshimaInteractiveSpeechSource2021,liHBPEfficientBlock2022,liFastMVAE2ImprovingAccelerating2023}.
Figure \ref{fig:perm} illustrates such a failure by an example of signals separated by IVA, where $M=N=2$ and each of the two source signals is colored either green or red.
As in Fig. \ref{fig:perm} (b), three frequency blocks appeared in this case: the lower- and higher-frequency blocks extracting the green source and the middle-frequency block for the red source. 
Consequently, IVA did not keep consistent permutations among these frequency blocks, resulting in poor separation performance.

\begin{figure}[t]
    \centering
    \includegraphics[width=\linewidth]{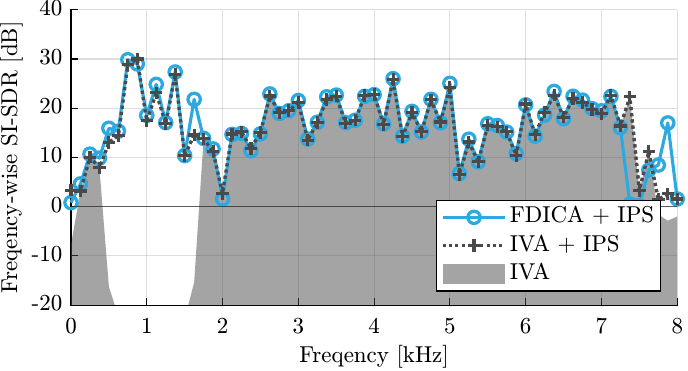}
    \caption{Frequency-wise SI-SDR of the separation result in Fig. \ref {fig:perm} (b), where the frequency bins are decimated by a factor of 16 for visibility.
    The light gray area represents the performance of IVA, where the block permutation problem arises in frequency bands from 0.5 to 1.5 kHz.
    The dotted line refers to IVA after the permutations are corrected using IPS. 
    For reference, the blue line shows the result of FDICA + IPS.
    }
    \label{fig:freqWiseDeltaSDR}
\end{figure}

However, even when the block permutation problem arises, the demixing matrices obtained for each frequency block are often optimized correctly, except for their permutations.
As an example, Fig. \ref{fig:freqWiseDeltaSDR} displays the frequency-wise separation performance of the result shown in Fig. \ref{fig:perm} (b). 
We calculated the frequency-wise version of the scale-invariant signal-to-distortion ratio (SI-SDR) of the 
separated signals $\mathbf{y}$ in the time-frequency domain as follows:
\begin{align}
    &\hspace{-7pt}\textrm{SI-SDR}_f \nonumber\\ 
    &\quad{}= 10 \log_{10} \left(\frac{
    \sum_{n=1}^N\sum_{t=1}^T|\tilde{s}_{ftn}|^2}{\sum_{n=1}^N\sum_{t=1}^T|\tilde{s}_{ftn}-y_{ftn}|^2}\right)\!,
\end{align}
where $\tilde{s}_{ftn}\in\mathbb{C}$ is the scaled version of the $n$th true source signal at the $(f,t)$th bin, i.e., 
\begin{align}
    \tilde{s}_{ftn}&= \left(\frac{\sum_{t=1}^T \bar{y}_{ftn} s_{ftn}}{\sum_{t=1}^T|s_{ftn}|^2}\right)\cdot s_{ftn},
\end{align}
and complex conjugation is denoted by $\bar{(\cdot)}$.
The area filled with light gray in Fig. \ref{fig:freqWiseDeltaSDR} corresponds to the separation result of IVA in Fig. \ref{fig:perm} (b).
The dotted line shows IVA + IPS, where IPS ideally resolved the permutation problem of the result of IVA so that the correlation between the separated signal and the true source is maximized (see Eq. \eqref{eq:bestpermutation}).
For reference, the blue line indicates FDICA with IPS (FDICA + IPS).
In this example, the separation performance of IVA was extremely bad within 0.5 to 1.5 kHz, which corresponds to the middle-frequency block in Fig. \ref{fig:perm} (b), where the red source is extracted.
However, IVA + IPS was comparable to that of FDICA + IPS across all frequencies, which highlights that if the block permutation was circumvented, then IVA performs effectively.
This observation motivates us to propose some specialized techniques to avoid the block permutation problem for IVA or other existing BSS algorithms.

While the block permutation problem arises from various factors, one major factor is considered to be the difficulty in handling the complicated structure of audio signals.
In speech signals, for example, there are two main components: vowels dominating the lower-frequency bands and consonants appearing in the higher-frequency bands.
When a BSS algorithm attempts to separate the sources,
it must align the permutations of these two components,
which is not straightforward since they appear in different frequency bands and at different times.
At the same time,  locally looking at a narrower frequency band (e.g., the mid-frequency block in Fig. \ref{fig:perm} (b)), only a few components are dominant and have a simple structure.
Therefore, it should be easy for existing BSS algorithms to resolve permutations inside such narrower bands.
If we split the BSS problem into a set of several BSS problems in the narrower frequency bands, 
then we can focus on how to align the permutations between the multiple bands.
The block permutation solvers can be used for this approach with some additional computational costs, but we can resolve the block permutation problem without an additional cost, as described in the next section.
    
\section{Proposed Method}
\label{sec:prop}
To solve the block permutation problem without paying additional costs,
we propose a simple technique named \textit{subband splitting}.
The proposed method splits all the frequencies into several subbands and sequentially separates them using existing BSS methods.
Narrowing the frequency band makes it easier to resolve permutations within each subband.
Then, the permutations between the subbands are aligned by using the separation result in one subband as the initial values for the subsequent subbands.
As in Fig. \ref{fig:perm} (c), our method enhances the separation performance of IVA without modifying the algorithm.
Here, we introduce the proposed technique and 
describe the way to generate the subbands used in our technique.
Subsequently, we note the number of iterations and 
the computational cost of the proposed technique.

\subsection{Subbands and BSS Method}

In the proposed method, the entire frequencies are split into $I$ subbands $(\mathcal{F}_i)_{i=1}^I$, where the $i$th subband $\mathcal{F}_i$ is the set of frequency indices, 
\begin{equation}
    \mathcal{F}_i = \left\{f\in \{1,\ldots,F\}\mid L_i \leq f \leq H_i\right\},
    \label{eq:subbands}
\end{equation}
$F$ is the number of frequency bins, and $L_i$ and $H_i$ $(L_i\leq H_i)$ are the lower and upper bounds for the $i$th subband $\mathcal{F}_i$, respectively.
All frequency indices must be contained in at least one subband, and hence $(\mathcal{F}_i)_{i=1}^I$ must satisfy
\begin{equation}
    \bigcup_{i=1}^I \mathcal{F}_i = \{1,\ldots, F\}.
\end{equation}
Additionally, the adjacent subbands must overlap, i.e.,
\begin{equation}
    \mathcal{F}_i\cap\mathcal{F}_{i+1} \neq \emptyset\qquad (1\leq i \leq I-1).
\end{equation}
For notational convenience, let the separation procedure of a BSS algorithm (e.g., IVA and ILRMA) be written as
\begin{equation}
(\textbf{y}, \mathbf{W},\boldsymbol{\Theta}) \leftarrow \mathsf{BSS}(\mathbf{x},\mathbf{W}^{(\text{init})},\boldsymbol{\Theta}^{(\text{init})}),
\end{equation}
where the BSS algorithm receives an observed signal $\mathbf{x}$, an initial value of the demixing matrix $\mathbf{W}^{(\text{init})}$, and initial values of other auxiliary variables $\boldsymbol{\Theta}^{(\text{init})}$ (e.g., NMF variables $(\mathbf{T}_n)_{n=1}^N$ and $(\mathbf{V}_n)_{n=1}^N$ of ILRMA in Eq.~\eqref{eq:costILRMA} or any other variables necessary for the BSS algorithm).
After running the BSS algorithm, it returns separated signals $\mathbf{y}$, the corresponding demixing matrix $\mathbf{W}$, and the corresponding auxiliary variables $\boldsymbol{\Theta}$.

\subsection{Proposed Method: Subband Splitting}

Using the above notation, the proposed subband splitting for any BSS algorithm $\mathsf{BSS}(\cdot)$ is given as follows:
\begin{equation}
\hspace{-10pt}
\left\lfloor\begin{aligned}\;
(\mathbf{x}_{\mathcal{F}_i},\mathbf{W}_{\!\mathcal{F}_i},\boldsymbol{\Theta}_{\mathcal{F}_i}) &\leftarrow  \mathsf{extract}_{\mathcal{F}_i}(\mathbf{x}, \mathbf{W},\boldsymbol{\Theta}),\\
(\mathbf{y}_{\mathcal{F}_i}, \mathbf{W}_{\!\mathcal{F}_i},\boldsymbol{\Theta}_{\mathcal{F}_i}) &\leftarrow \mathsf{BSS}(\mathbf{x}_{\mathcal{F}_i},\mathbf{W}_{\!\mathcal{F}_{i}},\boldsymbol{\Theta}_{\mathcal{F}_{i}}),\\
(\mathbf{y}, \mathbf{W},\boldsymbol{\Theta}) &\leftarrow  \mathsf{substitute}_{\mathcal{F}_i}(\mathbf{y}_{\mathcal{F}_i}, \mathbf{W}_{\!\mathcal{F}_i},\boldsymbol{\Theta}_{\mathcal{F}_i}),\vspace{-10pt}
\end{aligned}\right.
\label{eq:defSS}
\end{equation}
where $\mathsf{extract}_{\mathcal{F}_i}(\cdot)$ is the operator that extracts the part of variables corresponding to the $i$th 
subband $\mathcal{F}_i$.
The extracted part is indicated by the subscript $(\cdot)_{\mathcal{F}_i}$ as
\begin{align}
\mathbf{x}_{\mathcal{F}_i} = ((\mathbf{x}_{ft})_{f\in {\mathcal{F}_i}})_{t = 1}^{T},
\qquad
\mathbf{W}_{\!\mathcal{F}_i} = (\mathbf{W}_{\!f})_{f\in {\mathcal{F}_i}}.
\label{eq:ext}
\end{align}
The operator $\mathsf{substitute}_{\mathcal{F}_i}(\cdot)$ substitutes the extracted part of the variables back into their original location as 
\begin{align}
 ((\mathbf{y}_{\!ft})_{f\in {\mathcal{F}_i}})_{t = 1}^{T}\leftarrow\mathbf{y}_{\mathcal{F}_i},
\qquad
(\mathbf{W}_{\!f})_{f\in {\mathcal{F}_i}}\leftarrow \mathbf{W}_{\!\mathcal{F}_i},
\label{eq:sub}
\end{align}
namely, the part of optimization variables associated with the subband $\mathcal{F}_i$ are replaced with the latest optimization result provided by $\mathsf{BSS}(\cdot)$.
The proposed method repeats Eq.~\eqref{eq:defSS} for $i=1,\ldots,I$ so that $\mathsf{BSS}(\cdot)$ is applied to all the subbands.

Note that $\mathsf{extract}_{\mathcal{F}_i}(\cdot)$ and $\mathsf{subtitute}_{\mathcal{F}_i}(\cdot)$ for the auxiliary variables $\boldsymbol{\Theta}$ should be defined similarly, depending on the BSS algorithms used in each subband.
Specific definitions of the extraction/substitution operators for SS-IVA and SS-ILRMA are described in Section \ref{sec:deSSBSS}

The most important aspect of the proposed method is the existence of the overlap of the subbands.
In the overlapping part $\mathcal{F}_{i}\cap\mathcal{F}_{i+1}$,
the output of $\mathsf{BSS}(\cdot)$ within the $i$th subband is carried over to the $(i+1)$th subband as the initial value for $\mathsf{BSS}(\cdot)$.
Therefore, within the $(i+1)$th subband, the BSS algorithm is induced to obtain the same permutation as that obtained in the $i$th subband.
This initialization strategy can keep consistent permutation among the subbands if there are sufficient overlaps.

\subsection{Subband Generation by Shift Rules}
\label{sec:gensubband}

To generate the subbands, we define \textit{shift rules} that constantly shift the subbands upward or downward:
\begin{equation}
(L_{i+1}, H_{i+1}) = \begin{cases}
    (L_{i} + \Delta, H_i + \Delta) & (\textrm{Up})\\
    (L_{i} - \Delta, H_i - \Delta)  & (\textrm{Down})
\end{cases}
\label{eq:shift}
\end{equation}
where $\Delta>0$ is the amount of shift.

To easily compare the settings for subbands, we introduce \textit{subband parameter} $(\theta_W, \theta_\Delta)$.
The parameter $\theta_W\geq 1$ determines the width of subband $W$ $(=H_i-L_i+1)$
and the parameter $\theta_\Delta\geq 1$ controls the amount of shift $\Delta$ as follows:
\begin{align}
    W =\lceil F/\theta_W\rceil,
    \qquad
    \Delta= \lceil W/\theta_\Delta \rceil \approx F/(\theta_W\theta_\Delta),
    \label{params}
\end{align}
where $\lceil\cdot\rceil$ is the ceiling function.
That is, the width of each subband $W$ is $1/\theta_W$ times the entire number of frequency bins $F$.
The subband is shifted by the amount of another $1/\theta_\Delta$ times the bandwidth $W$.

To ensure that all the frequencies, including the highest and the lowest, are updated by the same number, we set the first bounds as follows:
\begin{equation}
\hspace{-10pt}(L_{1}, H_{1}) = \begin{cases}
    (\Delta-W+1,\Delta) & (\textrm{Up})\\
    (F-\Delta+1,F-\Delta+W)  & (\textrm{Down}).
\end{cases}
\label{eq:initLH}
\end{equation}
Using Eq. \eqref{eq:initLH}, all the frequency bins are included in $\theta_\Delta$ subbands.
Namely, for every frequency (including the lowest and the highest), BSS methods are applied by the same number, i.e., $\theta_\Delta$ times\footnote{
This may not be true when $\theta_\Delta$ is neither an integer nor a divisor of $W$.
For example, when $\theta_\Delta = 2.5$, BSS methods are applied either 2 $( {}=\lfloor \theta_\Delta \rfloor )$ or 3 $({} =\lceil \theta_\Delta \rceil )$ times for each frequency bin, where $\lfloor\cdot\rfloor$ denotes the flooring function.
}.
Note that the first and the last bounds overflow the entire frequency band $\{1,\ldots,F\}$, i.e., $L_i < 1$ or $H_i > F$ holds for smaller and larger $i$.
Such bounds are clipped to the range of $1$ to $F$ when each subband $\mathcal{F}_i$ is calculated (see Eq. \eqref{eq:subbands}),
and then the corresponding subbands become narrower than $W$. 
For instance, the first subband $\mathcal{F}_1$ and (when $\theta_W \theta_\Delta$ is a divisor of $F$) the last subband $\mathcal{F}_I$ have only $\Delta$ frequency bins.

\begin{figure}[t]
    \centering
    \includegraphics[width=\linewidth]{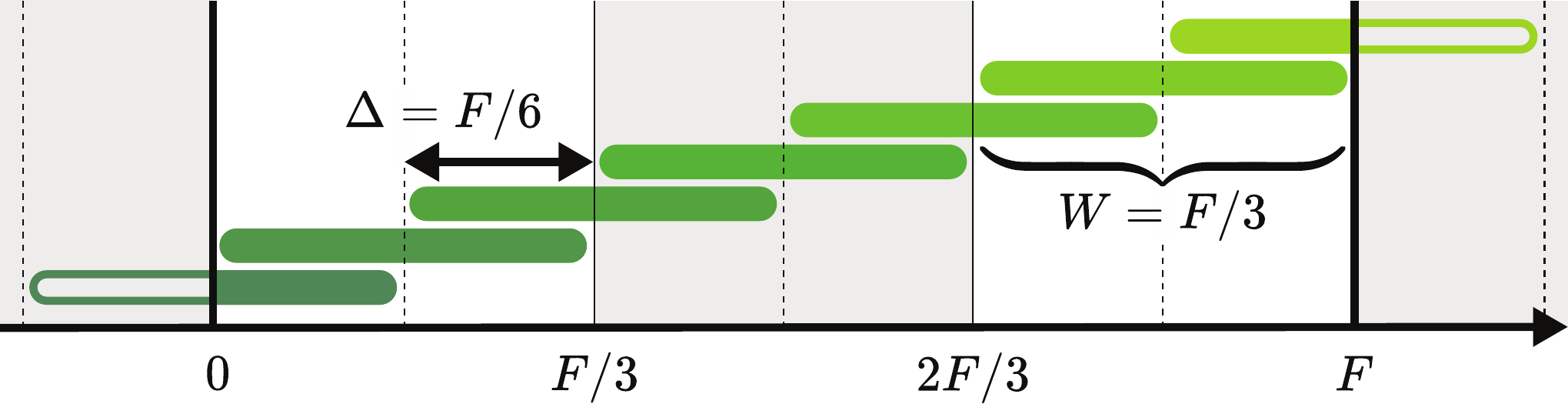}
    
\caption{Subbands generated by Eqs.~\eqref{eq:shift}, \eqref{params}, and \eqref{eq:initLH}.
The horizontal axis indicates the frequency index $f= 1,\ldots, F$. 
The bounds $(L_i, H_i)$ are indicated with the rounded edges of the bars, and its corresponding subband $\mathcal{F}_i$ is shown by their filled area. The subband parameter was set to $(\theta_W,\theta_\Delta) = (3,2)$.
Note that all the frequency bins are included in $2$ $({}=\theta_\Delta)$ subbands.}
\label{fig:subbandExample}
\end{figure}

Figure \ref{fig:subbandExample} illustrates an example of the subbands generated by the above rule.
Here, the subband parameter is set to $(\theta_W,\theta_\Delta) = (3,2)$, 
i.e., the bandwidth is $W=F/3$, and the amount of shift is $\Delta=F/6$.
All the frequency bins are included in 2 $({}=\theta_\Delta)$ subbands.
The subband splitting with the shift rule is summarized in Alg. \ref {alg:seq}, where any BSS method can be used for $\mathsf{BSS}(\cdot)$.

\begin{algorithm}[t]
\small
  \KwInput{$\mathbf{x}$, $\mathsf{BSS}(\cdot)$, $\theta_W$, $\theta_\Delta$}
  Initialize $\mathbf{W}, \boldsymbol{\Theta}$.\\
  Set $\Delta$ using Eq. \eqref{params}, and set $(L_1,H_1)$ using Eq. \eqref{eq:initLH}.\\
  $\mathcal{F}_{1} \gets \left\{f\in \{1,\ldots,F\}\mid L_{1} \leq f \leq H_{1}\right\}$\\
  $i \gets 1$\\
  \While{$\mathcal{F}_i \neq \emptyset$}
  {
    $(\mathbf{x}_{\mathcal{F}_i},\mathbf{W}_{\!\mathcal{F}_i},\boldsymbol{\Theta}_{\mathcal{F}_i}) \leftarrow  \mathsf{extract}_{\mathcal{F}_i}(\mathbf{x}, \mathbf{W},\boldsymbol{\Theta})$\\
    $(\mathbf{y}_{\mathcal{F}_i},\mathbf{W}_{\!\mathcal{F}_i},\boldsymbol{\Theta}_{\mathcal{F}_i}) \leftarrow \mathsf{BSS}(\mathbf{x}_{\mathcal{F}_i},\mathbf{W}_{\!\mathcal{F}_{i}},\boldsymbol{\Theta}_{\mathcal{F}_{i}})$\\
    $(\mathbf{y}, \mathbf{W},\boldsymbol{\Theta}) \leftarrow  \mathsf{substitute}_{\mathcal{F}_i}(\mathbf{y}_{\mathcal{F}_i},\mathbf{W}_{\!\mathcal{F}_i},\boldsymbol{\Theta}_{\mathcal{F}_i})$\\
    \textrm{Obtain next bound $(L_{i+1}, H_{i+1})$ by Eq. \eqref{eq:shift}}\\
    $\mathcal{F}_{i+1} \gets \left\{f\in \{1,\ldots,F\}\mid L_{i+1} \leq f \leq H_{i+1}\right\}$\\
    $i\leftarrow i + 1$
  }
  \KwOutput{$\mathbf{y}$}
\caption{Subband splitting with shift rules}
\label{alg:seq}
\end{algorithm}

\subsection{Number of Inner Iteration and Computational Cost}
Two iterations appear in the proposed method: 
(i) the \textit{outer iteration} $i=1,\ldots,I$, 
where $\mathsf{BSS}(\cdot)$ is sequentially applied to the subbands $(\mathcal{F}_i)_{i=1}^I$, and 
(ii) the \textit{inner iteration} that corresponds to the iterative updates in $\mathsf{BSS}(\cdot)$.

Here, we confirm the relationship between the parameter $\theta_\Delta$, the number of inner iterations $J_\textrm{inner}$, 
and the total number of updates $J_\textrm{total}$, 
i.e., how many times the demixing matrix $\mathbf{W}_f$ is updated for each frequency \footnote{
The total number of updates of the demixing matrix for each frequency will be the same because every frequency bin is included in the same number of subbands, as described in Section \ref{sec:gensubband}.
}.
Since the proposed method runs BSS algorithms $\theta_\Delta$ times duplicated for each frequency, the total number of updates increases as $\theta_\Delta$ increases, i.e., 
\begin{equation}
J_\textrm{total} = \theta_\Delta J_\textrm{inner}.
\end{equation}
For example, in Fig. \ref{fig:subbandExample}, all the frequency bins are updated twice because $\theta_\Delta=2$.

With fixing the number of inner iterations $J_\mathrm{inner}$,
the overall computational cost becomes 
$\theta_\Delta$ times larger due to the duplicated run of the BSS algorithm.
This increase in computational cost can be canceled
by dividing the number of inner iterations $J_\mathrm{inner}$ by $\theta_\Delta$.
Namely, given a total number of updates $J_\textrm{total}$, set the number of inner iterations as follows\footnote{
The increase in computational cost due to the duplicated runs of BSS algorithms is canceled using Eq. \eqref{eq:innerIter} when the cost of BSS algorithms depends linearly on the number of frequency bins.
This condition is satisfied with AuxIVA\cite{onoStableFastUpdate2011} and ILRMA\cite{kitamuraDeterminedBlindSource2016}.
}:
\begin{equation}
    J_\textrm{inner} = \lceil J_\textrm{total} / \theta_\Delta \rceil.
    \label{eq:innerIter}
\end{equation}
Note that the proposed method can improve memory efficiency because each subband is narrower than usual.

\section{Experiment}
\label{sec:exp}
In this section, we propose SS-IVA and SS-ILRMA by combining our subband splitting with IVA and ILRMA, respectively.
We also describe a conventional subband-aware method, which we call overlapped-clique-based IVA (OC-IVA) here. 
We then evaluate their separation performance and computational costs using speech signals.
We also evaluate the separation performance of SS-ILRMA using music signals.

\subsection{Proposed Method: SS-IVA and SS-ILRMA}
\label{sec:deSSBSS}

\subsubsection{SS-IVA}
The proposed SS-IVA uses AuxIVA\cite{onoStableFastUpdate2011} for the function $\mathsf{BSS}(\cdot)$ in Alg. \ref {alg:seq}.
Since $\mathbf{W}$ is the only variable optimized in AuxIVA, the auxiliary variable $\boldsymbol{\Theta}$ was omitted.

\subsubsection{SS-ILRMA}
The proposed SS-ILRMA uses ILRMA\cite{kitamuraDeterminedBlindSource2016} for $\mathsf{BSS}(\cdot)$  in Alg. \ref {alg:seq}.
Note that ILRMA has auxiliary variables $\boldsymbol{\Theta}=((\mathbf{T}_n)_{n=1}^N,(\mathbf{V}_n)_{n=1}^N)$ for NMF, i.e., the basis matrices $(\mathbf{T}_n)_{n=1}^N$ and the activation matrices $(\mathbf{V}_n)_{n=1}^N$.
Therefore, the operation of $\mathsf{extract}_{\mathcal{F}_i}(\cdot)$ and $\mathsf{substitute}_{\mathcal{F}_i}(\cdot)$ for $\boldsymbol{\Theta}$ must be defined, as well as $\mathbf{x}$, $\mathbf{W}$ and $\mathbf{y}$ in Eqs. \eqref{eq:ext} and \eqref{eq:sub}.
Here, we define the extraction operation for $\boldsymbol{\Theta}$ as follows:
\begin{align}
\boldsymbol{\Theta}_{\mathcal{F}_i} = (\mathbf{T}_{\mathcal{F}_i},(\mathbf{V}_n)_{n=1}^N),
\end{align}
where
\begin{equation}
    \mathbf{T}_{\mathcal{F}_i} = ((\mathbf{t}_{fn}^\mathsf{T})_{f\in\mathcal{F}_i})_{n=1}^N,
\end{equation}
and substitution operation
to do the reverse.
Namely, it extracts the part of basis matrices $(\mathbf{T}_n)_{n=1}^N$ associated with the subband $\mathcal{F}_i$, similar to demixing matrices $\mathbf{W}$. 
At the same time, it carries over the activation matrices $(\mathbf{V}_n)_{n=1}^N$ optimized in the $i$th subband to the $(i+1)$th subband as initial values.
This definition is based on the following intuition
that the activation $(\mathbf{V}_n)_{n=1}^N$ roughly reflects whether the $n$th source is active at each time, and they become similar in two adjacent subbands.
Carrying over a common activation matrices $(\mathbf{V}_n)_{n=1}^N$ provides a stronger bond between the subbands and helps avoiding block permutation problems.

\begin{figure*}[t]
    \centering
    \includegraphics[scale = 0.7]{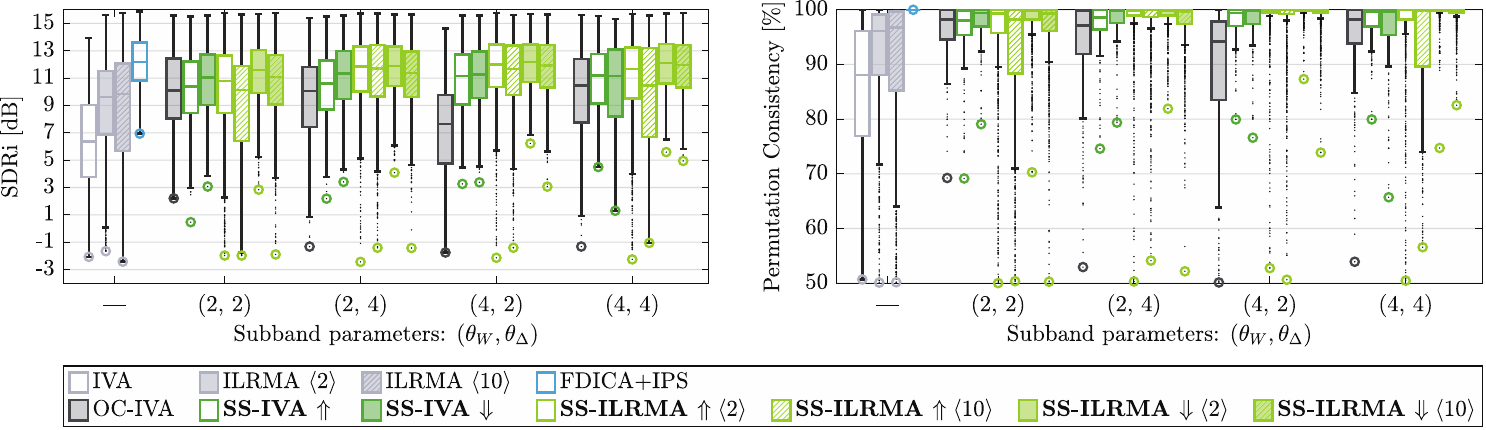}
    \caption{SDRi and permutation consistency of each method for speech signals.
    The proposed methods are emphasized by (light) green and bold letters.
    The number of basis for ILRMA $K$ is indicated by $\langle 2 \rangle$ and $\langle 10 \rangle$.
    Boxes for IVA-based methods contain 224 ($=$ 56 pairs of speech sources $\times$ 4 pairs of source directions) results, while those for ILRMA-based methods have 1120 (${}=224 \times 5$ seeds) results. Large markers show the worst cases. For the proposed SS-IVA and SS-ILRMA, the direction of the shift is indicated by $\Uparrow$ (upward) and $\Downarrow$ (downward).
    }
    \label{fig:exp1_perf}
\end{figure*}

\subsection{Comparison Method: OC-IVA\cite{leeIndependentVectorAnalysis2012}}

To compare our subband splitting with conventional subband-aware methods, 
we tested OC-IVA, which is an extension of IVA\cite{leeIndependentVectorAnalysis2012}.
Similar to the proposed method, 
OC-IVA considers overlapping subbands to estimate demixing matrices.
The significant difference with our SS-IVA is that OC-IVA separates all frequency bands simultaneously using an algorithm derived from the following objective function:
\begin{align}
&\mathcal{L}(\mathbf{W})=\sum_{n=1}^N\sum_{t=1}^T \sum_{i=1}^I\sqrt{\sum_{f\in\mathcal{F}_i}|y_{ftn}|^2}
\nonumber \\ &\quad\quad\quad\quad\quad\quad\quad\quad\quad
- 2T\sum_{f=1}^F\log(|\det(\mathbf{W}_f)|),
\end{align}
where we generate the subbands $(\mathcal{F}_i)_{i=1}^I$ in the same way with our method.
Note that the first term models the subband-wise group structure (see Eq. \eqref{eq:costIVA} for comparison to the original IVA).
Comparison with OC-IVA helps to see whether optimizing the subbands simultaneously or sequentially yields better performance.

\subsection{Experimental Settings}
\label{sec:commonCond}

We evaluated BSS methods using two-channel mixtures of two sources, i.e., $N=M=2$.
The speech and music signals were obtained from
SiSEC 2011\cite{araki2011SignalSeparation2012} dataset.
For speech signals (Section \ref{sec:exp1} and \ref{sec:exp2}), 8 speech signals were obtained from \texttt{dev1} in \textit{underdetermined speech and music mixtures} task\cite{araki2011SignalSeparation2012}.
Then, all 56 possible pairs (considering their permutations) were used as source signals.
Their duration was 10\,s, and the sampling frequency was 16\,kHz.
For music signals (Section \ref{sec:exp3}), we used 4 songs from \textit{professionally produced music recordings} task\cite{araki2011SignalSeparation2012}.
For each song, two instruments (including vocals, guitars, violins, and synth) were chosen as in \cite[Table 3]{kitamuraDeterminedBlindSource2016} (the paper proposed ILRMA).
They were downsampled to 16\,kHz.

The observed signals were generated by convolving room impulse responses in \cite{hadadMultichannelAudioDatabase2014} with downsampling to 16 kHz.
The reverberation time was 160\,ms.
The pair of source directions were $(-45^\circ, 30^\circ)$, $ (-75^\circ, 30^\circ)$, $(-45^\circ, 60^\circ)$ and $(-75^\circ, 60^\circ)$.
The spacing between the two microphones was 8\,cm, and the distance between the sources and the center of the microphones was 1\,m.

The demixing matrices $\mathbf{W}_{\!f}$ were initialized with identity matrices.
All elements of the initial value for $(\mathbf{T}_n)_{n=1}^N$ 
and $(\mathbf{V}_n)_{n=1}^N$ were drawn from the uniform distribution $\mathcal{U}(0,1)$, where five different seeds were used.

The window length and the hop size of STFT were set to 2048 and 1024 samples, respectively. 
The Hann window was used for the window function.
The Nyquist frequency was omitted from the optimization target, i.e., the number of frequency bins $F$ was $1024$.

The source-to-distortion ratio improvement (SDRi) \cite{vincentPerformanceMeasurementBlind2006} was used to evaluate separation performance.
In addition, we define \textit{permutation consistency}, a weighted accuracy of permutation, as follows:
\begin{equation}
\begin{aligned}
    &\text{Permutation Consistency} ={}\\ 
    &\quad\quad\quad\quad\quad\quad\max_{1\leq p\leq N!}\left(\frac{\textstyle\sum_{f=1}^F\gamma_f\cdot\delta_{p,q_f}}{\textstyle\sum_{f=1}^F\gamma_f}\right)\times 100,    
\end{aligned}
\end{equation}
where $\gamma_f$ is the frequency-wise weight given by the power of each frequency, i.e., 
\begin{equation}
    \gamma_f = \frac{1}{NT}\sum_{n=1}^N\sum_{t=1}^T|s_{ftn}|^2,
\end{equation}
$\delta_{ij}$ is Kronecker's delta, 
\begin{equation}
    \delta_{ij} = \begin{cases}
        1 & (i=j)\\
        0 & (i\neq j).
    \end{cases}
\end{equation}
$q_f$ is the index of correct permutation that achieves the highest correlation between the source and separated signal at each frequency, i.e., 
\begin{align} 
q_f &= \underset{1\leq q^\prime \leq N!}
{\mathrm{arg~max}}~
\sum_{n=1}^N\sum_{t=1}^T|s_{ftn}\,y_{ft(\sigma_{q^\prime n})}|,
\label{eq:bestpermutation}
\end{align}
where $\sigma_{qn}\in\{1,\ldots,N\}$, $\boldsymbol{\sigma}_q = (\sigma_{q1},\ldots,\sigma_{qN})\in S_N$ is the $q$th  permutation (i.e., $\{\boldsymbol\sigma_1,\ldots,\boldsymbol\sigma_{N!}\}=S_N$), and $S_N$ is the set of all permutations of $N$ elements.
Permutation consistency takes the values between $1/(N!)$ and $1$, and the higher, the better.
The experiments were performed with MATLAB R2024b on AMD Ryzen 9 5950X (3.40 GHz, 16 core).

\begin{table*}[t]
    \centering
    \tabcolsep = 4.7pt
    \caption{Average SDRi and permutation consistency for speech mixture. 
    Each column corresponds to a subband parameter $(\theta_W, \theta_\Delta)$, and the best value is bolded. 
    For the proposed SS-IVA and SS-ILRMA, the directions are indicated by $\Uparrow$ (upward) and $\Downarrow$ (downward).
    }
    \begin{tabular}{llcclccccccccccc}
    \hline
        \multicolumn{3}{c}{{\multirow{2}{*}{Method ($\star$: ours)}}} & {\multirow{2}{*}{$K$}}& 
         &\multicolumn{5}{c}{SDRi [dB]} & & \multicolumn{5}{c}{Permutation consistency [\%]}\\ \cline{6-10}\cline{12-16}
        & & & &  &{---} & $(2,2)$ & $(2,4)$ & $(4,2)$ & $(4,4)$ & & {---} & $(2,2)$ & $(2,4)$ & $(4,2)$ & $(4,4)$ \\
    \hline
IVA & Vanilla & \cite{onoStableFastUpdate2011} &  {---}&
 &{6.11} & {---} & {---} & {---} & {---} & & 85.42& {---}& {---}& {---}& {---}
\\
    & OC  & \cite{leeIndependentVectorAnalysis2012}& {---}&
 &{---} & {10.05} & {9.13} & {6.94} & {9.76} & & {---}& 96.16& 92.97& 88.3& 94.69
\\
& SS $\Uparrow$ & $\star$ & {---}&
 &{---} & {10.09} & {10.34} & {10.82} & {10.90} & & {---}& 96.25& 96.81& 97.55& 97.77
\\
& SS $\Downarrow$ & $\star$& {---}&
 &{---} & {10.73} & {11.00} & {10.96} & {10.53} & & {---}& 97.29& 97.65& 97.60& 95.98
\\
\hline
ILRMA  & Vanilla & \cite{kitamuraDeterminedBlindSource2016} & 2&
 &\bf{8.90} & {---} & {---} & {---} & {---} & & \bf{91.71}& {---}& {---}& {---}& 
{---}
\\
& Vanilla & \cite{kitamuraDeterminedBlindSource2016} & 10&
 &8.56& {---} & {---} & {---} & {---} & & 89.91& {---}& {---}& {---}& 
{---}
\\
& SS $\Uparrow$ &$\star$ & 2&
 &{---} & {10.01} & {11.29} & {11.50} & {10.82} & & {---}& 94.79& 97.75& 98.09& 96.46
\\
 & SS $\Uparrow$ & $\star$ & 10&
 &{---} & 8.91& 11.06& 11.03& 9.84& & {---}& 92.02& 96.90& 96.93& 93.97
\\
 & SS $\Downarrow$  & $\star$& 2& 
 &{---} & \bf{11.25} & \bf{11.67} & \bf{12.06} & \bf{11.95}&  & {---}& \bf{97.91}& \bf{98.78}& \bf{99.60}& \bf{99.49}\\ 
 &   SS $\Downarrow$  & $\star$& 10& 
 &{---} & 10.57& 10.97& 11.71& 11.79&  & {---}& 96.15& 97.21& 98.86& 99.12
\\ \hline
\multicolumn{3}{l}{\textcolor{gray}{FDICA + IPS}} & {---} &
 &\textcolor{gray}{12.20}& {---} & {---} & {---}  &  {---} & &  \textcolor{gray}{100.00}& {---}& {---}& {---}& {---}
\\
    \hline
    \end{tabular}
    \label{tab:table}
\end{table*}

\subsection{Separation Performance for Speech Signals}
\label{sec:exp1}

\begin{figure}[t]
    \centering
    \includegraphics[width=\linewidth]{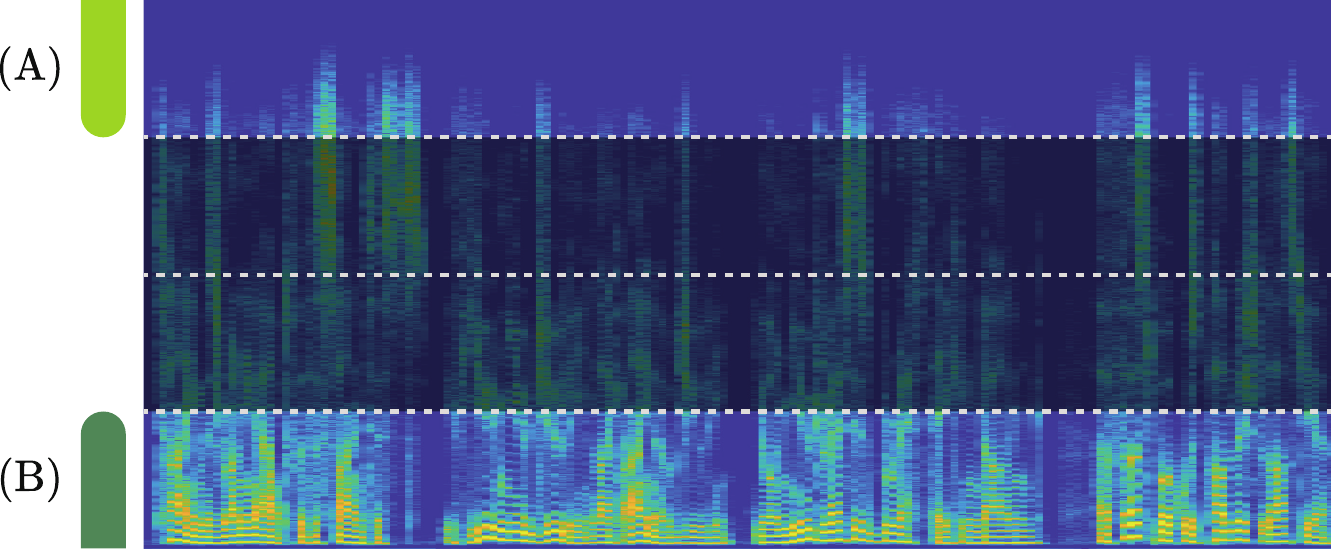}
    \caption{Example of a speech signal. 
    The signal is obtained from \texttt{dev1\_male4\_src\_2.wav} in \cite{araki2011SignalSeparation2012}.
    The vertical axis ranges from 0 to 8 kHz, and 
    the highest subband (A) and lowest subband (B) are highlighted.
    The subband parameters are set to $(\theta_W,\theta_\Delta)=(2,2)$.
    The middle-frequency band is darkened to emphasize the highest and lowest subbands.}
    \label{fig:example}
\end{figure}

We evaluated the separation performance of BSS methods using speech signals.
The total number of updates $J_\textrm{total}$ was fixed to 100.
The parameters $\theta_W$ and $\theta_\Delta$ were set to 2 or 4 to ensure that $\theta_\Delta\in\{2,4\}$ is a divisor of 100 (= $J_\textrm{total}$) and that $\theta_W \theta_\Delta\in\{2,4,8\}$ is a divisor of 1024 (= $F$).
The number of bases for ILRMA $K$ was set to 2 and 10.
We also tested FDICA + IPS, the oracle method that individually separates each frequency via FDICA and solves the permutation via IPS, where IPS finds the ideal permutation using Eq. \eqref{eq:bestpermutation}.

Figure \ref {fig:exp1_perf} and Table \ref{tab:table} summarize their separation performance, where \textit{Vanilla} refers to the original IVA or ILRMA handling all frequencies simultaneously. 
The experimental results show the notable improvements brought by the proposed subband splitting.

Both the conventional OC-IVA and our proposed SS-IVA outperformed Vanilla IVA, confirming that band splitting enhances separation performance. 
In all the settings, our SS-IVA (sequential separation)  consistently outperformed OC-IVA (simultaneous separation). 
Notably, the permutation consistency of SS-IVA was significantly high compared to OC-IVA, which demonstrates that the sequential procedure contributes to aligning permutation.

For ILRMA, the proposed SS-ILRMA obtained the highest performance for almost all subband parameters.
Downward SS-ILRMA with $(K,\theta_W,\theta_\Delta)=(2,4,2)$ achieved the best result in the experiment.
Remarkably, its separation performance was comparable to FDICA + IPS, and the permutation consistency was almost perfect (99.60 on average).
Moreover, even under the worst case (indicated by a large marker), it resulted in $6.22$ dB in SDRi, which was $7.88$ dB higher than Vanilla ILRMA ($-1.66$ dB).
Note that the average performance of ILRMA and SS-ILRMA tended to be better when $K=2$ rather than when $K=10$.
At the same time, when $(\theta_W,\theta_\Delta)=(4,4)$, the downward SS-ILRMA with $K=10$ achieved similar performance as $K=2$ (see Fig. \ref{fig:exp1_perf}), indicating that splitting into smaller subbands reduces the size of the problem within each subband and helps the algorithms to work stably.

The downward shift tended to be superior to the upward shift, which was particularly noticeable in SS-ILRMA.
That is, starting with the separation of higher frequency bands was more effective.
This is likely due to the sparsity (an important cue for many BSS algorithms) of the speech signals in higher frequency bands.
To illustrate this, Fig. \ref{fig:example} shows a speech signal used in the experiment along with the (A) highest and (B) lowest subbands, where the subband parameter was set to $(\theta_W,\theta_\Delta)=(2,2)$.
When using the upward shift, the lowest band (B) must be separated first.
However, this subband is relatively dense, making it difficult for sparsity-based BSS methods.
Failures at this subband negatively affect the separation of subsequent subbands.
On the other hand, when the downward shift is used, the highest subband (A) is separated first.
Owing to its sparsity and simple structure (i.e., components concentrated in specific time frames and appearing as vertical patterns), this subband can be successfully separated using IVA and ILRMA.
The results from the higher subbands are then propagated to the lower (i.e., more challenging) subbands and help their separation, leading to better outcomes.

\subsection{Computational Costs}
\label{sec:exp2}

\begin{figure}[t]
    \centering
    \includegraphics[scale = 0.7]{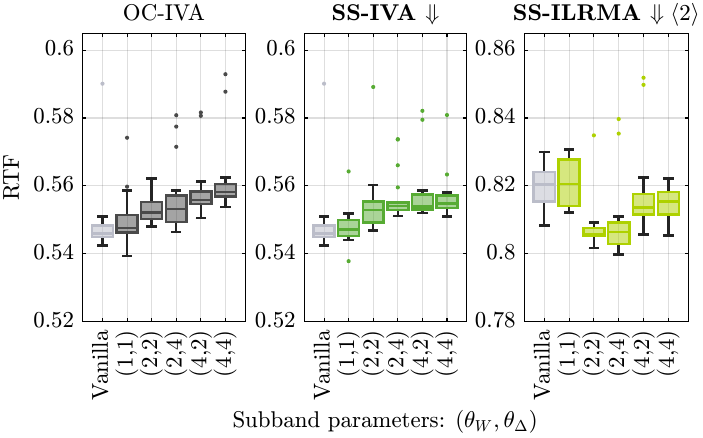}
    \caption{RTF with various subband parameters.
    The proposed methods are emphasized by (light) green and bold letters. 
    The downward shift was used for SS-IVA and SS-ILRMA.
    The number of basis $K$ for ILMRA is set to $2$, as indicated by $\langle 2 \rangle$.}
    \label{fig:exp2_RTF}
\end{figure}

\begin{figure}[t]
    \centering
    \includegraphics[scale = 0.7]{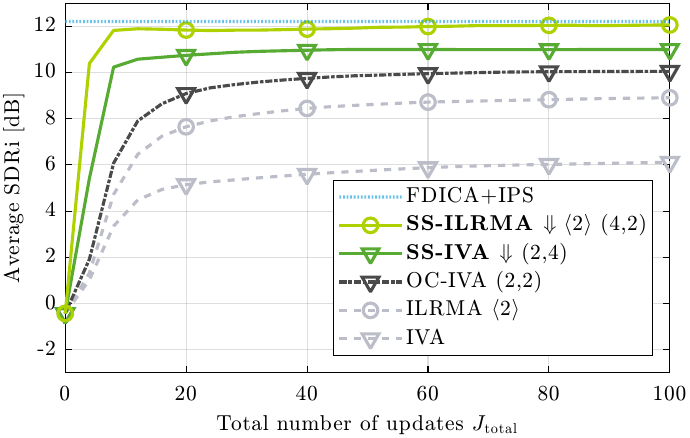}\\
    \subcaption{Per number of updates.}\\
    \vspace{\baselineskip}
    \includegraphics[scale = 0.7]{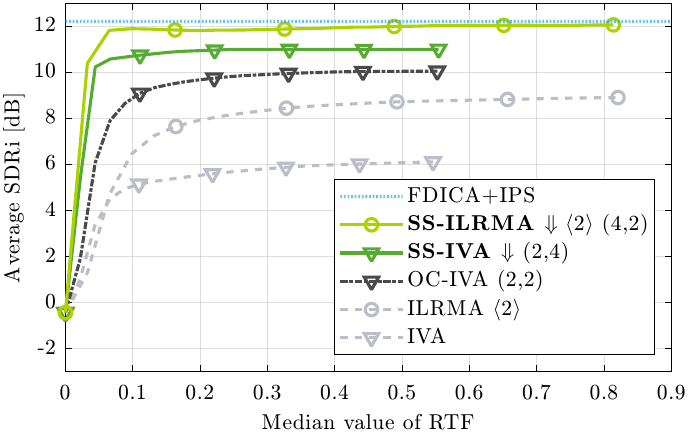}
    \subcaption{Per RTF.}
    \caption{Transition of average SDRi per (a) total number of updates $J_\mathrm{total}$ and (b) RTF for speech signals. 
    Markers are shown for each 20 updates.
    The proposed methods are emphasized by (light) green and bold letters.
    The downward shift was used for SS-IVA and SS-ILRMA.
    The number of basis $K$ for ILMRA and SS-ILRMA was set to $2$, as indicated by $\langle 2 \rangle$.
    The best subband parameters $(\theta_W,\theta_\Delta)$ in Table \ref{tab:table} was used for each method.
    The separation performance of FDICA+IPS is shown by the horizontal line.}
    \label{fig:exp2_per_time}
\end{figure}

Next, we investigate the computational costs of the proposed technique using the same mixtures as in Section \ref{sec:exp1}
We run each BSS method 20 times to measure the real-time factor (RTF), i.e., the normalized runtime required to process one second of the signal.
To check the implementations, we also tested SS-IVA and SS-ILRMA with $(\theta_W,\theta_\Delta)=(1,1)$, which agrees with the Vanilla version of IVA and ILRMA.
The total number of updates $J_\textrm{total}$ was fixed to 100.
The downward shift was used, and the number of bases $K$ was set to $2$ since SS-IVA and SS-ILRMA were found to perform better with this setting (see Section \ref{sec:exp1}).

The results are shown in Fig. \ref{fig:exp2_RTF}. 
The conventional OC-IVA and the proposed SS-IVA took slightly longer runtime than Vanilla IVA due to band splitting. 
Their RTF increased as the subband parameters $\theta_W$ and $\theta_\Delta$ grew.
For ILRMA, the subband splitting was found to reduce the computation time.
It was also observed that the runtime in SS-ILRMA was primarily influenced by $\theta_W$, which determines the bandwidth $W$, and was less dependent on $\theta_\Delta$, which controls the amount of shift $\Delta$.

We also examined the relationship between computational time and separation performance.
The total number of updates $J_\textrm{total}$ was varied from 0 to 100 in increments of 4.
For each method, the best subband parameters $(\theta_W,\theta_\Delta)$ in Table \ref{tab:table} were used: $(2,2)$ for OC-IVA, $(2,4)$ for downward SS-IVA, and $(4,2)$ for downward SS-ILRMA with $K=2$.

Figure \ref{fig:exp2_per_time} (a) shows the average separation performance against the total number of updates $J_\textrm{total}$.
The separation performance of conventional IVA and ILRMA grew gradually, which is due to some mixtures that require many iterations for resolving the permutation problem.
On the other hand, the proposed SS-IVA and SS-ILRMA rapidly converged in fewer iterations.
SS-IVA reached the ceiling of the separation performance around $J_\textrm{total} = 12$ $({}=\theta_\Delta \times J_\textrm{inner} = 4 \times 3)$, 
and SS-ILRMA around $J_\textrm{total} = 8$ $({}=\theta_\Delta \times J_\textrm{inner} =  2\times 4)$.
The required number of updates $J_\textrm{total}$ (i.e., 12 and 8) is less than half compared to that in the previous literature, e.g., \cite{kitamuraDeterminedBlindSource2016}, thanks to the ability to avoid erroneous permutation.

Figure \ref{fig:exp2_per_time} (b) shows the average separation performance versus RTF.
The proposed SS-IVA and SS-ILRMA were confirmed to converge with shorter runtimes than conventional methods.

\begin{figure}[t]
    \centering
    \includegraphics[scale = 0.7]{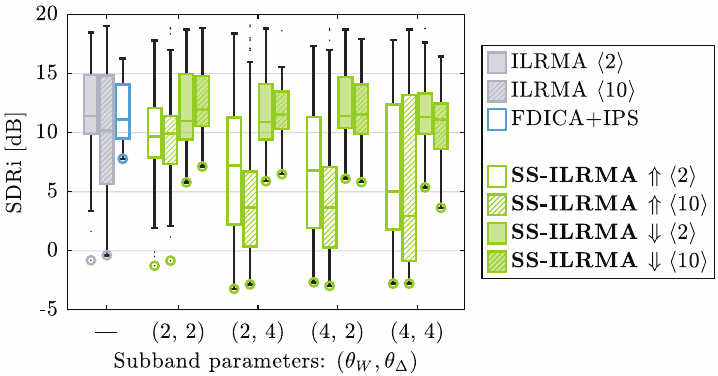}\\
    \caption{SDRi of each method for music signals.
    The proposed methods are emphasized by (light) green and bold letters.
    The number of basis for ILRMA $K$ is indicated by $\langle 2 \rangle$ and $\langle 10 \rangle$.
    Each box contains 80 ($=$ 4 pairs of music sources $\times$ 4 pairs of source directions $\times$ 5 seeds) results. 
    Large markers show the worst cases. For the proposed SS-ILRMA, the direction of the shift is indicated by $\Uparrow$ (upward) and $\Downarrow$ (downward).}
    \label{fig:exp3_music}
\end{figure}

\subsection{Separation Performance for Music Signals}
\label{sec:exp3}
From the experimental results so far, we have confirmed that, for speech signals, SS-ILRMA is highly effective. 
This might be a surprising outcome because ILRMA is known to be unsuitable for speech signals but
suitable for music signals.
Therefore, we additionally evaluated SS-ILRMA on music signals.
The parameters are the same as in Section \ref{sec:exp1}

Figure \ref{fig:exp3_music} and Table \ref{tab:tableMusic} show the separation performance of ILRMA, SS-ILRMA, and FDICA+IPS.
When the number of basis $K$ was set to 10, the Vanilla ILRMA performed worse in some cases\footnote{
When only evaluating the song ID4 in \cite[Table 3]{kitamuraDeterminedBlindSource2016} (i.e., guitar and synth in \cite[Fig. 12]{kitamuraDeterminedBlindSource2016}
that investigates the relationship between $K$ and SDRi), the average SDRi of Vanilla ILRMA was 16.24 dB ($K=2$) and 16.62 dB ($K=10$), i.e., $K=10$ outperformed $K=2$, which agrees with \cite[Fig. 12]{kitamuraDeterminedBlindSource2016}.
The poor performance of Vanilla ILRMA with $K=10$ in Fig. \ref{fig:exp3_music} is due to the other mixtures, especially ID2 and ID3.
}.
The upward SS-ILRMA was ineffective for music signals, while the downward SS-ILRMA performed robustly for both $K=2$ and $K=10$. 
In particular, the downward SS-ILRMA with $(K,\theta_W, \theta_\Delta) = (10,2,2)$ yielded the best results in terms of the median SDRi, where the worst-case SDRi was improved by 7.51 dB compared to Vanilla ILRMA with $K=10$.
Note that downward SS-ILRMA with $(K,\theta_W, \theta_\Delta) = (2,2,2)$, $(2,4,2)$, and $(10,2,2)$ even outperformed FDICA+IPS on average, which might be because SS-ILRMA can more precisely model the structure of the power spectrogram (including the relationship among the time-frequency bins) by using NMF, while FDICA handles the frequency bins independently.

\begin{table}[t]
    \centering
    \tabcolsep = 2.6pt
    \caption{Average SDRi of ILRMA (Vanilla), SS-ILRMA (SS) and FDICA+IPS for music signals. 
    Each column corresponds to a subband parameter $(\theta_W, \theta_\Delta)$, and the best value is bolded. 
    For the proposed SS-ILRMA, the directions are indicated by $\Uparrow$ (upward) and $\Downarrow$ (downward).
    }
    \begin{tabular}{lcccccccc}
    \hline
        \multicolumn{2}{c}{Method} & {\multirow{2}{*}{$K$}} & & 
        \multicolumn{5}{c}{SDRi [dB]} \\ \cline{5-9}
        \multicolumn{2}{c}{($\star$: ours)} &  & & {---} & $(2,2)$ & $(2,4)$ & $(4,2)$ & $(4,4)$ \\
    \hline
Vanilla &  \cite{kitamuraDeterminedBlindSource2016} & 2 & & \bf{11.65}
 & {---} & {---} & {---} & {---}\\
Vanilla & \cite{kitamuraDeterminedBlindSource2016} & 10 & & {9.99} & {---} & {---} & {---} & {---}\\
SS $\Uparrow$  & $\star$ & 2& & {---} & {9.75}  & {7.02}& {6.96} & {6.97} \\
SS $\Uparrow$ & $\star$  & 10 &  & {---} & {9.75} & {4.67} & {4.66} & {5.75}\\
SS $\Downarrow$ & $\star$& 2& & {---} & {11.89} & {11.48} & \bf{12.32} & \bf{11.59} \\ 
SS $\Downarrow$ &  $\star$ & 10 & & {---} & \bf{12.29} & \bf{11.53} & {11.61} & {10.85} \\ \hline
\multicolumn{2}{l}{\textcolor{gray}{FDICA+IPS}} & {---} & & \textcolor{gray}{11.71} & {---} & {---} & {---} & {---} \\ \hline
    \end{tabular}
    \label{tab:tableMusic}
\end{table}

\begin{figure}[t]
    \centering
    \includegraphics[width=\linewidth]{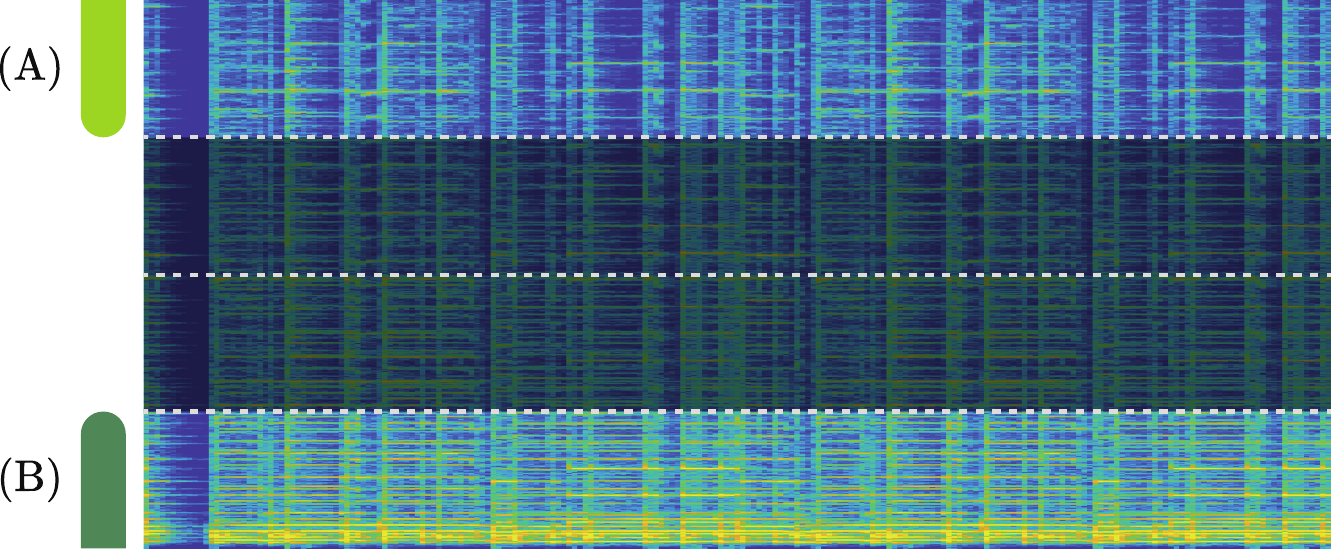}
    \caption{Example of a music signal (guitar).
    The signal is obtained from \texttt{dev1\_\_bearlin-roads\_\_snip\_85\_99}
    \texttt{\_\_acoustic\_guit\_main.wav} in \cite{araki2011SignalSeparation2012}.
    The vertical axis ranges from 0 to 8 kHz, and 
    the highest subband (A) and lowest subband (B) are highlighted.
    The subband parameters are set to $(\theta_W,\theta_\Delta)=(2,2)$.
    The middle-frequency band is darkened to emphasize the highest and lowest subbands.}
    \label{fig:exampleMusic}
\end{figure}

The difference between upward and downward shifts was more pronounced in music signals compared to that in speech signals (Section \ref{sec:exp1}).
This should be because the separation of lower frequency bands in music signals can be significantly challenging for some instruments. 
As an example, Fig. \ref{fig:exampleMusic} shows the signal of the guitar used in the experiment. 
Similar to speech signals, the highest subband (A) is sparse.
On the contrary, the lowest subband (B) is even denser than that of speech signals.
Therefore, the downward shift, i.e., leveraging the separation result obtained from higher subbands to assist in the separation of lower subbands, is indispensable when using SS-ILRMA for music signals.

\section{Conclusion}
\label{sec:conclusion}

In this paper, we proposed \textit{subband splitting}, a technique for improving the performance of existing BSS algorithms.
It sequentially applies a BSS algorithm to several overlapping subbands. Despite its simplicity, the proposed method effectively improves the separation performance.
Additionally, we proposed SS-IVA and SS-ILRMA by incorporating IVA and ILRMA into our technique. 
Experimental results showed that our downward SS-ILRMA reached significantly higher separation performance with rapid convergence.
Future work includes integrating more advanced BSS methods (e.g., those utilizing deep learning) for further improvement of separation performance. 
It is also valuable to extend the proposed technique to real-time processing scenarios by leveraging its fast convergence.

\bibliographystyle{jasj}
\bibliography{Bibliography}

\profile{Kazuki Matsumoto}{is currently an undergraduate student at Waseda University. His research interests include optimization-based signal processing.}

\profile{Kohei Yatabe}{received the B.E., M.E., and Ph.D. degrees from Waseda University in 2012, 2014, and 2017, respectively. He is currently an Associate Professor at the Department of Electrical Engineering and Computer Science at the Tokyo University of Agriculture and Technology.}

\end{document}